\documentclass[graybox]{svmult}

\usepackage{mathptmx}    % selects Times Roman as basic font
\usepackage{helvet}   % selects Helvetica as sans-serif font
\usepackage{courier}  % selects Courier as typewriter font
\usepackage{type1cm}  % activate if the above 3 fonts are
\usepackage{makeidx}   % allows index generation
\usepackage{graphicx}  % standard LaTeX graphics tool
\usepackage{multicol}  % used for the two-column index
\usepackage[bottom]{footmisc}% places footnotes at page bottom
\usepackage{aas_macros}
\usepackage{url}
\makeindex    % used for the subject index
\begin{document}

\title*{The Cavendish \textit{Computors}: The women working in scientific computing for Radio Astronomy}
\titlerunning{The Cavendish \textit{Computors}}
\author{Verity Allan}
\institute{Verity Allan \at Battcock Centre for Experimental Astrophysics, Cavendish Laboratory, University of Cambridge \email{vla22@mrao.cam.ac.uk}}

\maketitle

\abstract*{A discussion of the women doing work related to scientific computing for Radio Astronomy in the Cavendish Laboratory in the decades after the Second World War.} 

\abstract{A discussion of the history of scientific computing for Radio Astronomy in the Cavendish Laboratory of the University of Cambridge in the decades after the Second World War. 
This covers the development of the aperture synthesis technique for Radio Astronomy and how that required using the new computing technology developed by the University's Mathematical Laboratory: the EDSAC, EDSAC 2 and Titan computers. 
This article looks at the scientific advances made by the Radio Astronomy group, particularly the assembling of evidence which contradicted the Steady State hypothesis. 
It also examines the software advances that allowed bigger telescopes to be built: the Fast Fourier Transform (FFT) and the degridding algorithm. 
Throughout, the contribution of women is uncovered, from the diagrams they drew for scientific publications, through programming and operating computers, to writing scientific papers.}
% Always give a unique label
% and use \ref{<label>} for cross-references
% and \cite{<label>} for bibliographic references
% use \sectionmark{}
% to alter or adjust the section heading in the running head
%\begin{quotation}
%Please do not use quotation marks when quoting texts! Simply use the \verb|quotation| environment -- it will automatically render Springer's preferred layout.
%\end{quotation}
\section{Introduction}

The years following the Second World War were a time of massive scientific and technological innovation in the University of Cambridge. 
The University's Mathematical Laboratory was set up prior to the War to investigate numerical methods, and after the war, Maurice Wilkes became its director.
Having visited the United States and seen the ENIAC (Electronic Numerical Integrator and Computer) computer, he started building EDSAC (Electronic Delay Storage Automatic Calculator), putting Cambridge at the forefront of the computer revolution \cite[p.~41]{Ahmed}.
In the same post-War years, the Cavendish Radio Astronomy group assembled, with Martin Ryle at its head. The Cavendish Laboratory of the University of Cambridge had been doing fundamental physics research for decades, and staff had won Nobel Prizes \cite{Cavendish-Laboratory:2021}.
The Radio Astronomy group would add to this collection, as Ryle and Tony Hewish won the Nobel Prize for Physics in 1974 for their innovative telescope design and the discovery of pulsars --- work that relied on the computers of the Mathematical Laboratory \cite{Nobel}.

This reliance on computers is exemplified by Ryle's request for a bigger computer:

\begin{quotation}
One day, Ryle came to me to say that he was planning the erection of a much larger telescope and to ask whether the Mathematical Laboratory could undertake to provide the computing support required. \cite[p.~193]{Wilkes}
\end{quotation}

This paper will explain why Radio Astronomy needed a bigger computer, and will elucidate the role of other members of the Radio Astronomy group and the Mathematical Laboratory beyond the relatively well-known names of Ryle, Hewish, and Wilkes.

It has become more widely known in recent years that a lot of early computing relied on women --- thanks to work by Janet Abbate and Mar Hicks and Margo Lee Shetterly \cite{Abbate}, \cite{Hicks}, \cite{Shetterly}. 
Women were clearly involved in the early years of computing at Cambridge, as can be seen from the archives of the Department of Computer Science and Technology, which is a successor to the Mathematical Laboratory. 
A photograph of staff from 1949 shows 20 staff, of whom 8 were women \cite{Computer-Laboratory}.
In this paper, we will uncover some of the contributions of women to science and scientific programming in Radio Astronomy in the years from 1950 to 1963.

The University computers that the Radio Astronomy group used during this period were EDSAC, the first electronic digital stored program computer in the world, EDSAC 2, which led to many innovations in microprogramming, and Titan, which led to advances in computer graphics and computer aided design (CAD) \cite[p.~72]{Ahmed}. Titan was the computer requested by Ryle, which was needed for the One-Mile telescope.

\section{The Women In Radio Astronomy Computing}

The Radio Astronomy group became reliant on computing, and by 1965, over 40\% of all papers produced by the group had required the use of a computer to process data \cite[p.~14]{Leedham}.
Women were extensively involved in this endeavour. Around 50\% of people acknowledged in papers for their contribution to the programming and operation of computers were women \cite[p.~14]{Leedham}.
These women were employed in a variety of roles.

Some were hired as \textit{Computors} in the 1950s and 1960s,\footnote{Source: Cavendish Laboratory Archive Personnel Records.} initially to do calculations, and then later to assist with programming. 
\textit{Computor} or \textit{Computer} as a job description was common in the 1940s, first to describe someone who was hired to do calculations by hand or with a mechanical or electro-mechanical computer, and then to describe a worker who used a computer in some capacity  \cite{Light}. 
The Cavendish records also show that people (often women) were hired as \textit{scanners} for High Energy Physics: their job was to analyse photographic plates from cloud chamber experiments. 
Women were also hired as Research Assistants: Elizabeth Waldram was hired as such \cite[p.~63]{Ahmed}. 
Judy Bailey, who later became Deputy Director of the University Computing Service (another successor to the Mathematical Laboratory) worked for Radio Astronomy as a Technical Officer \cite{University-of-Cambridge}.

Women were also hired to prepare diagrams for scientific papers, drawing graphs and contour diagrams, as it was not yet possible to prepare these with computers. 
While preparing this paper, because of the Covid-19 pandemic, it was not possible for me to look at University records to get a sense of the number of people in these roles, nor of how many of them were women. 
They were known as \textit{the girls in the attic} \cite{Clifford}, p. viii.

Some of the women involved in scientific computing were PhD students and researchers themselves. 
Ann Neville, one of Ryle's PhD students, did a considerable amount of programming. 
She had to program for EDSAC 2, and to save space, many constants had to be combined into one constant, just to shave a few bytes off the size of the program.\footnote{Donald Wilson, Personal Communication, 22 April 2021.}

\section{The first forays into scientific computing}

While Charles Wynn-Williams was a pioneer of electro-mechanical computing for the Cavendish  \cite{Wynn-Williams}, the Radio Astronomy group did not leap straight to digital computing for all their papers. 
We can trace the use of computers and the contributions of programmers primarily through the acknowledgements section of papers. 
They published their first paper using EDSAC in 1953, and use through the 1950s was fairly sporadic  \cite[p.~14]{Leedham}. 

The first women we encounter in the acknowledgements probably were the computers.
In 1951, a Mrs A. C. Hollis-Hallett is credited as having \textit{analysed a large proportion of the records} \cite[p.~963]{Smith}. 
It's not possible to verify this, as no records survive.
Similarly, a Mrs Moore is credited in a paper by Scheuer and Ryle from 1953, \textit{for help in the computation of results} \cite[p.~17]{Scheuer}. 
It's likely that Mrs Moore was the computer, not EDSAC, as EDSAC was not listed in the acknowledgements (which was the common practice at the time).

However, the development of a new technique in Radio Astronomy was going to change the group's relationship with computers: the technique of aperture synthesis.

The basic principles were sketched out by Ryle in a 1957 paper \cite{Ryle:1957}, and implemented by John Blythe. Aperture synthesis, as the name suggests, allows for the synthesis of a larger aperture. 
By using two or more smaller telescopes, one can achieve the resolution (but not the sensitivity) of a telescope with diameter equivalent to the longest spacing between antennas.\footnote{The distance between a given pair of antennas is known as a \textit{baseline}: the length of the longest baseline dictates the resolution achievable by the telescope.}
However, when measuring a signal with a particular pair of antennas, only part of the signal is observed -- the measurement is of a Fourier component of the signal \cite{Christiansen}, Chapter 8.
To reconstruct the complete signal, the other Fourier components must be measured, which can be achieved by moving the antennas around to fill in the aperture. 
The Fourier components then need to be combined and transformed into a useful measurement of sky brightness. 

Thus the Fourier Transform must be calculated for aperture synthesis telescopes, and this is not really feasible to do by hand. 
So the Radio Astronomy group was forced to turn to computers. 
Blythe's papers from 1957, where he implemented this technique, used EDSAC for the calculations \cite[p.~651]{Blythe:1957a}, \cite[p.~659]{Blythe:1957b}.
One of these papers used EDSAC to perform 15 hours of computation, performing over 2 million operations, to output data in a form suitable for making contour maps \cite{Blythe:1957a}. 
In both papers, Blythe credited a Mrs J. Sandos \textit{who performed most of the analysis}. 
No-one in the Radio Astronomy group can recall Mrs Sandos (who may have been a member of the Mathematical Laboratory), but she undoubtedly helped with the complex data preparation process: error-free punched tapes were required as input, and all the data had to be arranged in a particular way in order to compensate for EDSAC's lack of storage space. 

By 1960, we see EDSAC 2 appearing in paper acknowledgements. However, the workflow still retains a number of manual steps, as one paper notes: 

\begin{quotation}
It would be possible to record such information directly in digital form, but it has been found relatively easy to transfer the present records to punched tape in a record reader and manual perforator. \cite[p.412]{Costain}.
\end{quotation}

The acknowledgements are silent about who did this, but it's quite likely to be either PhD students or \textit{Computors}.

\section{Earth Rotation Aperture Synthesis and the Steady State Hypothesis}
In 1961, an extension of the aperture synthesis technique was developed: using the rotation of the Earth to help fill in the aperture of a telescope, thus performing Earth Rotation Aperture Synthesis. 
This technique was prototyped using the Cambridge 178MHz telescope and EDSAC 2 --- the new, more powerful, computer in the Mathematical Laboratory.
This telescope and others were used to collect data that seriously undermined the Steady State hypothesis.

In 1961, Scott, Ryle and Hewish performed earth-rotation aperture synthesis for the first time \cite{Scott:1961}. 
This paper has a number of interesting features. Firstly, this paragraph:
\begin{quotation}
The processes of filtering and convolution are basically equivalent and in this case where the output of the receiver was already recorded in digital form it was convenient to carry out a direct convolution with a time function whose Fourier transform provides the required polar diagram. \cite[p.~100]{Scott:1961}.
\end{quotation}
This is an early use of a digital filter rather than a piece of electronics; the work is moving from hardware to software. Note also that the \textit{digital form} is paper tape!

To record the data, a \textit{sample and hold} electronic circuit was implemented electro-mechanically, using photo-electric cells, to output two columns of 5-hole tape for each data point (giving ten bits of precision).\footnote{Thanks to B.J. Harris for spotting that they were building a sample and hold circuit.}
Each tape started at the same sidereal time: at the same time with respect to the fixed stars in the sky, which is most useful to astronomers, so non-local astronomical bodies can be located easily. 
The sidereal time arrangement meant that the data could be efficiently processed --- while EDSAC 2 had more storage space than EDSAC, programmers still could not afford to be profligate. During the data input phase, the data were convolved with a function which allowed for error correction, and which also allowed the extraction of the sine and cosine components which were required for the main computation. 
At this point, the data were sorted into a more convenient order before being written onto magnetic tape.
This magnetic tape was used as the input data for the main calculation, in which the data were put through the Fourier Transform. 
The final outputs were effectively a 2-dimensional map of the sky, sampled at discrete intervals.
The probability distribution of the measured intensity was also calculated and output.
The final values for the sky map needed to be interpolated, but this was done manually, possibly by the \textit{girls in the attic} or junior members of the group.

As well as being the first practical demonstration of aperture synthesis using the earth to fill in the aperture, this paper also was important for providing evidence against the Steady State hypothesis. The authors note: 

\begin{quotation}
there is no evidence for a reduction in the density of radio sources... this observation implies an isotropy which extends to at least 10,000 sources steradian. \cite[p.~111]{Scott:1961}.
\end{quotation}

This isotropy in extra-galactic faint sources conflicted with predictions from the Steady State model, which predicted that there should be many fewer faint sources, and that there should be no difference between galactic and extra-galactic sources. 
Once again we find a woman tucked away in the acknowledgements: Mrs R. Feinstein. Ruth Feinstein was employed by the Mathematical Laboratory  \cite{Computer-Laboratory:1999}, and was an assistant to Dr Lucy Slater, who became Head of Computing in the University's Department of Applied Economics  \cite{Andrews}, p.6. Ruth Feinstein assisted her husband, Charles, with \textit{calculating and typing} for his PhD thesis  \cite[p.~4]{Offer}. Unfortunately, she left the Mathematical Laboratory shortly after marriage  \cite{Computer-Laboratory:1999}, and she seems to play no further part in computing for Radio Astronomy. She was, along with Dr David Wheeler,responsible both for the running and design of the programme  \cite[p.~111]{Scott:1961}.\footnote{Note that at this time, the UK still used \textit{programme} to refer to computer programs.}

\subsection{The Steady State Hypothesis}
A summary of the steady state hypothesis is that the universe is expanding, but has no beginning and no end, and that its appearance does not change over time. 
In contrast, the Big Bang theory states that the universe expanded from an initial state of high density and temperature, and thus its appearance changes over time.

Ryle’s group was building up to attack the Steady State hypothesis.
There are two papers from Harriet Tumner and Patricia Leslie, both PhD researchers in the group \cite{Leslie, Tumner}. 
Neither paper mentioned using EDSAC 2, though both used data from the Cambridge surveys that were processed using that computer. 
These papers look at the brightness, density, size, and clustering of extra-galactic radio sources --- where the Steady State and Big Bang hypotheses generated different predictions.

Then there is a 1961 paper from Paul Scott and Martin Ryle  \cite{Scott:1961b}. This looks at the relationship between the number of radio sources, and their flux density (essentially their brightness. The Steady State hypothesis assumes that there is a particular relationship between the brightness of the source, and how many sources of that brightness there are.

The acknowledgements in this paper read: 
\begin{quotation}
We should like to thank Dr Hewish and Miss Patricia Leslie for useful discoveries, and Miss Ann Neville and Mr DR Marks for assistance in the reduction of observations. \cite[p.~397]{Scott:1961b}
\end{quotation}
Ann Neville was one of Ryle’s PhD students, and she will appear as an author on later papers.

Next there’s a paper by Hewish. 
He modelled the source number--flux density relationship, and used EDSAC 2 to do it. 
This is one of the earliest uses of the Monte Carlo method in Radio Astronomy: successors to this method are being used today as as key modelling technique  \cite{Feroz}.
Hewish doesn’t credit anyone else, so he most likely did a lot of the programming himself. 
This contrasts with Ryle, who didn’t program at all, even though Ryle was well aware of how essential computers were for aperture synthesis.\footnote{Dr Elizabeth Waldram, personal communication 4 May 2021.}

Finally, there is a Ryle and Clarke paper of 1961 in which they showed that extra-galactic radio sources have no variation in density in different parts of the sky --- the only variation comes when observing the Milky Way galaxy (which is nearby in astronomical terms \cite{Ryle:1961}.
The observed source number-flux density relationship doesn't match the steady state hypothesis. 
They conclude: 
\begin{quotation}
These observations do, however, appear to provide conclusive evidence against the steady state model.
\end{quotation}
Any last hopes that the Steady State theorists had of salvaging their model was blown away conclusively in 1964 when Penzias and Wilson detected the Cosmic Microwave Background \cite{Penzias}. 

\section{A radio survey of the North polar region}
We now turn to analyse in some detail a paper from 1962, in which major advances were made in both computing and Radio Astronomy \cite{Ryle:1962}. 

There are two major computational innovations in this paper. The first is Wheeler’s
Fast Fourier Transform (FFT) algorithm, tersely described in a single paragraph, plus
an equation. 
From this, we can glean  that the data from the sky needs
to be placed onto a Cartesian grid with $(x, y)$ co-ordinates. 
This means data must be converted from spherical co-ordinates that are most naturally used for observing objects on the sky into Cartesian co-ordinates.  
While the mathematician Gauss developed the theory of FFT, this practical implementation pre-dates the one by Cooley and Tukey in 1965, which is usually credited as the first practical implementation \cite{Heideman}.

The other programming innovation was the map degridding algorithm, which enables conversion back from Cartesian co-ordinates into spherical co-ordinates, so the radio sources are correctly located on the sky, which also ended up producing some of the world’s first computer graphics. 
The output from the
computer was plotted on a CRT (Cathode Ray Tube  monitor, where it lasted for a few seconds, just long
enough to be photographed \cite[p.~63]{Ahmed}.
This work was done by Elizabeth Waldram, then a Research Assistant in the group. This degridding algorithm (and its successors) were used for all subsequent interferometers built by the group.\footnote{Dr Elizabeth Waldram, personal communication 04 May 2021.} 

With the FFT and the degridding algorithm, it was possible to build a map of the northern radio sky.
It’s very likely that the final construction of the map from the photographic
plates was made by the \textit{girls in the attic}. 
This kind of mapping was a large part of their job. 
They weren't working directly with computers themselves, but they were a vital part of the scientific workflow, often alongside secretarial staff who would type up manuscripts. 
We can see a rare acknowledgment for the women doing such work in a 1962 paper  \cite{Turtle}: \begin{quotation}
We are grateful to Miss C. Dunn and Mrs M. Warren for helping with the analysis and the preparation of the contour maps. 
\end{quotation}
The contour maps were drawn from EDSAC 2 output \cite[p.~476]{Turtle}.

With this paper, we have the foundations for modern radio astronomy computing. 
The FFT saves a lot of computational effort compared to the Direct Fourier Transform which works with spherical co-ordinates. 
However, the use of the FFT also necessitates the use of gridding and degridding algorithms, which we need to first transform our data into the right shape for the FFT, and then put it back into spherical co-ordinates afterwards.
FFT and gridding algorithms remain an important part of Radio Astronomy research \cite{Ye}, \cite{Veenboer}, \cite{Nishimura}.

It's clear from Wheeler's inclusion in the acknowledgements rather than as a co-author that the exclusion of women programmers from authorship wasn't purely a result of sexism. 
Sexism may still have played a part, as a majority of those acknowledged for programming in Radio Astronomy papers were women \cite[p.~14]{Leedham}. 
Mar Hicks has shown how work done by women, especially in the computing field, was devalued, which may also have been a contributing factor \cite[passim]{Hicks}. 
Wheeler was acknowledged in Ryle's Nobel Prize lecture for his Fast Fourier Transform algorithm \cite{Ryle:1974}; none of the women get mentioned.

\section{Discussion}
It's clear from the preceding sections that women were deeply involved in the production of scientific papers in the 1950s and 60s, and that they were also involved in doing much of the programming and a good deal more of the science than is usually recognised. The programming also required good mathematical skills and an understanding of Radio Astronomy, so that programmers could work with sidereal time and Right Ascension and Declination (the spherical co-ordinate system used by astronomers) and their conversion into Cartesian co-ordinates for use in the FFT --- something that mathematics undergraduates often struggle with. Perhaps the \textit{girls in the attic} didn't require such advanced understanding, but they were expected to work to the highest standards of accuracy. 

It is extraordinarily difficulty to get a picture of many of the computational methods used. Programs were not routinely recorded at the time, though there are some contemporaneous program booklets stored in the archive of the Department of Computer Science and Technology.\footnote{See for example, shelfmark V75-58, which lists programs used by the Department on Geology on Titan.} However, none of these survive for Radio Astronomy. The information presented in papers is minimal --- usually just enough to allow for some mathematical recreation, but not enough to understand the algorithms. This may have been the result of the culture of the research group at the time; Ryle may have maintained more of a war-time culture of secrecy than we find desirable as scientists today.\footnote{Donald Wilson, personal communication 22 April 2021.}

It is also worth noting that, while women may have left the Mathematical Laboratory on marriage, this wasn't true for members of the Radio Astronomy group. Indeed, programmers were in such high demand that they could work part-time and flexibly, many years ahead of this becoming widely available. Dr Waldram recalls working 10-15 hours a week in the early 1960s, to accommodate her childcare needs.\footnote{Elizabeth Waldram, personal communication 4 May 2021.}

\section{Conclusion}
The acknowledgements to papers provide one of the few ways of finding out about the work done in early scientific computing, as they often indicate that work was done using a computer, and who did that work. We can get tantalising glimpses of what that work was, and the people behind it. This could usefully be supplemented by more archival work within the University, once this is possible again.

\begin{acknowledgement}
My thanks go to Dr Elizabeth Waldram, one of the Cavendish's computing pioneers, who has given her time very generously, to Dr Donald Wilson, who provided many valuable insights into this period, to Dr David Green, who helped me with my \LaTeX, to J. Nevins, who helped proof-read, and to Professor Malcolm Longair, who has supported this project.
\end{acknowledgement}

\bibliography{cavcomp}
\end{document}